\documentclass{amsart}
\usepackage{amssymb}
\usepackage{amsfonts}

\newtheorem{theorem}{Theorem}
\theoremstyle{plain}

\newtheorem{axiom}{Axiom}

\numberwithin{equation}{section}
\input{tcilatex}

\begin{document}
\title[Mechanics of complex bodies]{Mechanics of complex bodies: commentary
on the unified modelling of material substructures}
\author{Paolo Maria Mariano}
\address{DICeA, University of Florence,\\
via Santa Marta 3, I-50139 Firenze (Italy)}
\email{paolo.mariano@unifi.it}
\date{March 6$^{th}$, 2008}
\subjclass[2000]{Primary 05C38, 15A15; Secondary 05A15, 15A18}
\keywords{Complex bodies, material substructures, covariance}

\begin{abstract}
Basic issues of the general model-building framework of the mechanics of
complex bodies are discussed. Attention is focused on the representation of
the material elements, the conditions for the existence of ground states in
conservative setting and the interpretation of the nature of the various
balance laws occurring.
\end{abstract}

\maketitle
\tableofcontents

\section{Introduction}

Materials in which changes in the molecular or crystalline texture at
various microscopic scales (substructure) influence the macroscopic behavior
through peculiar interactions are commonly available. Liquid crystals,
ferroelectrics, quasicrystals, polymeric fluids are paradigmatic examples.
The attribute \emph{complex} is assigned to bodies made of these materials
in order to underline that significant substructural effects must be
accounted for.

In complex bodies the prototype material element is a \emph{system}. Often
it is a perfectly identifiable Lagrangian system, like in nematic liquid
crystals in which the characteristic stick molecules embedded in a soft
matrix may be extracted and isolated from the rest. Sometimes it does not
and the substructure is in a certain sense \emph{virtual}, like in
microcracked bodies: microcracks do not exist \emph{per se}, rather they are
determined only by the surrounding matter. On the other hand, the
substructure can be procedural in the sense that it is a consequence of
local rearrangements of the matter due to phase transitions from one
energetic well to another as in martensite-austenite mixtures.

Notwithstanding the variety of phenomena displayed by complex bodies and
classified in condensed matter physics, there exists an abstract
model-building framework for the mechanics of complex bodies. It unifies in
a unique format existing models of special classes of complex bodies and is
a flexible tool for analyzing new materials.

Basic aspects of such a model-building framework are discussed here with
pedagogical purposes and constructive criticism. Some appropriate references
are scattered throughout the subsequent sections.

\section{Representation of the morphology of material elements}

In its primitive meaning, a body can be regarded as an abstract set $%
\mathcal{B}$ collecting \emph{material elements}, each one being the
smallest piece of matter characterizing the material composing the body. The
basic issue is the `representation' of such a set. In the standard format of
continuum mechanics the geometrical representation adopted is the minimal
one: each material element is mapped onto a place that it occupies in the
ambient space $\mathbb{R}^{d}$. However, in the real world the material
elements are groups of entangled molecules, simple or complex pieces of
crystalline structures, stick molecules dispersed in a ground fluid etc.,
and substructural changes may generate actions that it is hard to identify
only with perturbations of the standard stress. In all these cases the
standard representation of bodies is too minimalist. The material element
should be considered in essence a \emph{system} rather than a windowless box
so that, in the representation of the body, a map attributing to each
material element a \emph{morphological descriptor} $\nu $ of its (inner)
substructure has to be defined. To construct the essential structures of the
mechanics of complex bodies, at least at the level of first principles, it
is not necessary to render precise the nature of the morphological
descriptor except assuming that it is an element of a finite-dimensional
differentiable manifold $\mathcal{M}$ with minimal geometrical properties
(each property of $\mathcal{M}$ has in fact physical meaning, see \cite{M02}%
). Once a reference place $\mathcal{B}\ $for the entire body is selected in $%
\mathbb{R}^{d}$, any other actual place $\mathcal{B}_{a}$ (with the index $a$
meaning actual) is considered to be achieved in an isomorphic copy $\hat{%
\mathbb{R}}^{d}$ of $\mathbb{R}^{d}$ itself by means of a \emph{%
transplacement} map 
\begin{equation*}
\mathcal{B}\ni x\mapsto y\left( x\right) \in \mathcal{B}_{a}
\end{equation*}%
which is assumed to be one-to-one, differentiable and orientation
preserving, with spatial derivative%
\begin{equation*}
F:=Du\left( x\right) \in Hom\left( T_{x}\mathcal{B},T_{u\left( x\right) }%
\mathcal{B}_{a}\right) .
\end{equation*}%
The geometry of the inner structure of the material elements is described
(at least at a coarse grained level) by a \emph{morphological descriptor map}
\begin{equation*}
\mathcal{B}\ni x\longmapsto \nu \left( x\right) \in \mathcal{M}
\end{equation*}%
which is assumed differentiable with spatial derivative 
\begin{equation*}
N:=D\nu \left( x\right) \in Hom\left( T_{x}\mathcal{B},T_{\nu \left(
x\right) }\mathcal{M}\right) .
\end{equation*}%
Motions are then time-parametrized families of transplacements and
morphological descriptors, namely for $x\in \mathcal{B}$ and $t\in \left[ 0,%
\bar{t}\right] \subset \mathbb{R}$ the time%
\begin{equation*}
\left( x,t\right) \longmapsto y:=y\left( x,t\right) \in \mathbb{\hat{R}}^{d},
\end{equation*}%
\begin{equation*}
\left( x,t\right) \longmapsto \nu :=\nu \left( x,t\right) \in \mathcal{M},
\end{equation*}%
both fields assumed to be piecewise twice differentiable in time so that%
\begin{equation*}
\dot{y}:=\frac{d}{dt}y\left( x,t\right) \ \text{ \ and\ \ \ \ }\dot{\nu}:=%
\frac{d}{dt}\nu \left( x,t\right)
\end{equation*}%
represent the macroscopic velocity and the rate of change of the
substructure respectively.

\section{Standard and substructural actions}

\subsection{Observers}

The description of material substructures constrains to give a detailed look
to the essence of the notion of observer. I have discussed the point
repeatedly (see e.g. \cite{M08} and references therein). The essential
aspects of the discussion are summarized below (with additional remarks) as
a preamble to the use of a weaker invariance requirement under changes in
observers presented later. The aim is to underline how the geometrical
features of the ambient space $\hat{\mathbb{R}}^{d}$\ and the manifold of
substructural shapes $\mathcal{M}$ influence the structure of the integral
balances of the interactions of macroscopic and substructural nature.

My point of view is that an observer should be considered as a
representation of all geometrical environments necessary to describe the
morphology of a given body and its motion. To this aim, in standard
continuum mechanics one needs to select the reference place $\mathcal{B}$,
the ambient space, say $\hat{\mathbb{R}}^{d}$, and the interval of time.
They are all the geometrical environments needed. Different is the stage
when substructural complexity arises and its changes influence the gross
behavior. The manifold of substructural shapes comes into play in the way
described above and one needs to represent it, specifically, one needs to
select atlantes over $\mathcal{M}$.

Once $\mathcal{M}$ is accounted for, the definition of changes in observers
should involve the representations of $\hat{\mathbb{R}}^{d}$, $\mathcal{B}$, 
$\left[ 0,\bar{t}\right] $, $\mathcal{M}$. In particular, the attention is
focused here on changes in observers leaving invariant $\mathcal{B}$ and the
interval of time. The last requirement defines synchronous changes in
observers. Really, one could consider affine time rescaling: non essential
consequences accrue so that they are not considered here.

\begin{itemize}
\item Changes in the ambient space are given by elements of $Diff(\hat{%
\mathbb{R}}^{d},\hat{\mathbb{R}}^{d})$, the group of diffeomorphisms of $%
\mathbb{\hat{R}}^{d}$ onto itself. Really one takes smooth curves 
\begin{equation*}
s\longmapsto \mathbf{f}_{s}\in Diff(\hat{\mathbb{R}}^{d},\hat{\mathbb{R}}%
^{d}),\text{ \ \ }s\in \mathbb{R}^{+},
\end{equation*}%
with $\mathbf{f}_{0}=id$, where $id$ means identity. The parameter $s$ can
be identified with the time so that the curve $s\longmapsto \mathbf{f}_{s}$
can be interpreted in the common way as the motion by which two observers
differ as the time flows. In particular the vector field $y\longmapsto
v\left( y\right) :=\frac{d\mathbf{f}_{s}}{ds}\left\vert _{s=0}\right. $ can
be considered as the (virtual) velocity of an observer moving with respect
to another one, a velocity pulled back in the frame of the first observer.

\item The material substructures are placed in the ambient space and the
manifold $\mathcal{M}$ collects only the elements of a concise description
of the characteristic features of their geometry in space. Thus changes of
frames in the ambient space alter in principle the geometry of the
substructures and their consequent representation over $\mathcal{M}$.
Disconnection between changes in the manifold of substructural shapes and
changes of frames in space is admissible only when $\nu $ represents only a
generic property of the substructures not associated with their geometry in
space. Changes in the choice of atlantes over $\mathcal{M}$ are governed by
elements of the Lie group of diffeomorphisms of $\mathcal{M}$ onto itself,
namely%
\begin{equation*}
G:=\left\{ g:\mathcal{M\longrightarrow M}\text{ }|\text{ }g\text{ a
diffeomorphism}\right\} .
\end{equation*}%
The link with changes in the ambient space are then assured by assuming the
existence of a homeomorphism%
\begin{equation*}
h:Diff(\hat{\mathbb{R}}^{d},\hat{\mathbb{R}}^{d})\rightarrow G,\text{ \ \ }%
h\left( id\right) =id_{G},
\end{equation*}%
where $id_{G}$ is the identity over $G$, so that a curve%
\begin{equation*}
s\longmapsto \nu _{g}:=h_{s}\left( \nu \right) ,
\end{equation*}%
with $h_{s}=h\left( \mathbf{f}_{s}\right) $. By indicating by $\xi $ the
element of the Lie algebra $\mathfrak{g}$ of $G$ given by the derivative $%
\frac{dg_{s}}{ds}\left\vert _{s=0}\right. =\frac{dh\left( \mathbf{f}%
_{s}\right) }{ds}\left\vert _{s=0}\right. $, its value over a given $\nu $
is indicated by $\xi _{\mathcal{M}}\left( \nu \right) :=\frac{d\nu _{s}}{ds}%
\left\vert _{s=0}\right. $. In particular, if the curve $s\longmapsto 
\mathbf{f}_{s}$ is selected over the special orthogonal group $SO\left(
d\right) $, a subgroup of $Diff(\hat{\mathbb{R}}^{d},\hat{\mathbb{R}}^{d})$,
for $\wedge q$ an element of the Lie algebra $\mathfrak{so}\left( d\right) $%
, $q\in \hat{\mathbb{R}}^{d}$, it is possible (and also convenient) to write 
$\xi _{\mathcal{M}}\left( \nu \right) $ as the product $\mathcal{A}\left(
\nu \right) q$ with $\mathcal{A}\left( \nu \right) \in Hom(\hat{\mathbb{R}}%
^{d},T_{\nu }\mathcal{M})$.
\end{itemize}

\emph{Isometric semi-classical changes in observers} are the ones that, by
leaving invariant $\mathcal{B}$ and $\left[ 0,\bar{t}\right] $, are
characterized by the choice of the subgroup of $Diff(\hat{\mathbb{R}}^{d},%
\hat{\mathbb{R}}^{d})$ coinciding with the semi-direct product $\hat{\mathbb{%
R}}^{d}\ltimes SO\left( d\right) $.

\emph{Rotational semi-classical changes in observers} are the ones in which
the subgroup of $Diff(\hat{\mathbb{R}}^{d},\hat{\mathbb{R}}^{d})$ selected
as ambient of $s\longmapsto \mathbf{f}_{s}$ is just $SO\left( d\right) $. In
this case, by indicating by $\dot{y}^{\ast }$ and $\dot{\nu}^{\ast }$ the
rates evaluated after the change in observer (they are the pull-back in the
frame of the first observer of the rates evaluated by the second observer),
one gets%
\begin{equation}
\dot{y}^{\ast }=\dot{y}+q\wedge \left( y-y_{0}\right) ,  \label{CO1}
\end{equation}%
where $y_{0}$ is an arbitrarily fixed centre of rotation in the ambient
space, and%
\begin{equation}
\dot{\nu}^{\ast }=\dot{\nu}+\mathcal{A}\left( \nu \right) q.  \label{CO2}
\end{equation}%
Remind that $q$ depends only on the parameter $s$ that is identified here
with the time, so that $q=q\left( t\right) $.

\subsection{Augmented external power and $SO\left( d\right) $ invariance}

A \emph{part} of $\mathcal{B}$ is any subset $\mathfrak{b}$ of $\mathcal{B}$
itself with non-vanishing volume measure and the same geometric `regularity'
of $\mathcal{B}$ (that is the same topological properties). All parts of $%
\mathcal{B}$ form an algebra $\mathfrak{P}\left( \mathcal{B}\right) $ with
respect to the operations of meet and join (see \cite{CV}).

Let $Vel$ be the space of all rate fields $\left( x,t\right) \longmapsto 
\dot{y}\left( x,t\right) $ and $\left( x,t\right) \longmapsto \dot{\nu}%
\left( x,t\right) $\ over the tube $\mathcal{B}\times \lbrack 0,\bar{t}]$,
rates calculated along possible motions $\left( x,t\right) \longmapsto
y\left( x,t\right) $ and $\left( x,t\right) \longmapsto \nu \left(
x,t\right) $.

A \emph{power} along motions $\left( y,\nu \right) $ is a functional%
\begin{equation*}
\mathcal{P}:\mathfrak{P}\left( \mathcal{B}\right) \times Vel\rightarrow 
\mathbb{R}
\end{equation*}%
which is \emph{additive over disjoint parts} and \emph{linear in the rates}.

The explicit representation of a given $\mathcal{P}$\ requires
classification of the interactions occurring in a body. The interest here is
on the expression of the power of all \emph{external} actions over a generic
part $\mathfrak{b}$ and along a motion $\left( y,\nu \right) $, a functional
indicated from now on by $\mathcal{P}_{\mathfrak{b}}^{ext}\left( \dot{y},%
\dot{\nu}\right) $.

Once $\mathfrak{b}$ is selected arbitrarily, interactions with the rest of
the power and the external environment are classified in two main
subclasses: (1) \emph{standard actions} power conjugated with the
macroscopic deformation, (2) \emph{substructural actions} associated with
the rate of change of the substructure inside the material elements. Each
subclass is further subdivided into bulk and contact actions which admit
densities with respect to volume and surface measures $dx$ and $d\mathcal{H}%
^{d-1}$, respectively.

The natural expression of the external power satisfying previous assumptions
is then given by%
\begin{equation*}
\mathcal{P}_{\mathfrak{b}}^{ext}\left( \dot{y},\dot{\nu}\right) :=\int_{%
\mathfrak{b}}\left( b\cdot \dot{y}+\beta \cdot \dot{\nu}\right) \text{ }%
dx+\int_{\partial \mathfrak{b}}\left( \mathbf{t}\cdot \dot{u}+\boldsymbol{%
\tau }\cdot \dot{\nu}\right) \text{ }d\mathcal{H}^{d-1},
\end{equation*}%
a power written in referential form. The term `augmented' in the title of
this section underlines the presence of the power densities of the
substructural actions. At any $x$ in $\mathfrak{b}$ one gets 
\begin{equation*}
b:=b\left( x\right) \in T_{y\left( x\right) }^{\ast }\mathcal{B}_{a}\simeq 
\mathbb{\hat{R}}^{d},\ \beta :=\beta \left( x\right) \in T_{\nu \left(
x\right) }^{\ast }\mathcal{M},
\end{equation*}%
while, for $x\in \partial \mathfrak{b}$,%
\begin{equation*}
\mathbf{t}:=\mathbf{t}\left( x\right) \in T_{y\left( x\right) }^{\ast }%
\mathcal{B}_{a}\simeq \mathbb{\hat{R}}^{d},\ \boldsymbol{\tau }:=\boldsymbol{%
\tau }\left( x\right) \in T_{\nu \left( x\right) }^{\ast }\mathcal{M}.
\end{equation*}%
Cauchy theorem indicates that the standard traction $\mathbf{t}$ can be
expressed in Lagrangian representation by means of the first Piola-Kirchhoff
stress tensor $P$, so that $\mathbf{t}=Pn$ at any $x\in \partial \mathfrak{b}
$ where $n$ is defined. $P\left( x\right) $ belongs to $Hom(\mathbb{R}%
^{d},T_{y\left( x\right) }^{\ast }\mathcal{B}_{a})$ and also it is natural
to consider it as the density of a form over $\mathcal{B}$. The argument of
the proof is the standard Cauchy's one based on the tetrahedron, or subtle
refinements of it (see the results in \cite{Sil}). In analogous way one may
presume that $\boldsymbol{\tau }=\mathcal{S}n$ with $\mathcal{S}\left(
x\right) \in Hom(\mathbb{R}^{d},T_{\nu \left( x\right) }^{\ast }\mathcal{M})$%
. Of course, even the microstress $\mathcal{S}$\ can be considered as the
density of a form over $\mathcal{B}$. The standard tetrahedron argument does
not apply here because the field $x\longmapsto \boldsymbol{\tau }\left(
x\right) $ takes values on the whole cotangent bundle $T^{\ast }\mathcal{M}%
=\cup _{\nu \in \mathcal{M}}T_{\nu }^{\ast }\mathcal{M}$ so that its total
over a generic side of the tetrahedron is not defined, as pointed out more
in general later. To prove the existence of $\mathcal{S}$\ by using the
tetrahedron argument, it is necessary to embed $\mathcal{M}$ in a linear
space (the relevant proof is in \cite{CV}). The embedding always exists
since $\mathcal{M}$ is assumed here to be finite-dimensional (Whitney
theorem) and can be selected to be also isometric (Nash theorem), this
choice having the advantage to preserve the quadratic part of the
independent kinetic energy that can be sometimes attributed to the material
substructure (see relevant comments in \cite{M08}). However, the theorems
indicating the availability of the embedding of $\mathcal{M}$ in a linear
space do not assure the uniqueness of the embedding itself. More precisely,
the embedding is neither unique nor rigid ($\mathcal{M}$ can be at the end
folded in various manners in the process). Additionally, the dimension of
the target linear space depends on the regularity of the embedding (Nash
theorem). In all cases, if one finds convenient embedding $\mathcal{M}$ in a
linear space for technical purposes, for example for constructing
appropriate finite elements for some special model of complex bodies, the
choice of the embedding becomes strictly a matter of modelling. Of course,
such a peculiarity disappears when the complex material under examination
admits a manifold of substructural shapes which is coincident with a linear
space, as in the case of micromorphic or, more generally, affine bodies (see 
\cite{Min}).

One might na\"{\i}vely claim that $\mathcal{M}$ is always a linear space so
that the scheme of affine bodies (the one discussed for example in \cite{Ger}%
, \cite{GR}) is sufficient for analyzing the material complexity at
substructural level.

Solids with distributed magnetic spins in conditions of magnetic saturation (%
$\mathcal{M}$ coincides with $S^{2}$) and the superfluid helium $^{4}He$
(where $\mathcal{M}=S^{1}\subset \mathbb{C}$) are elementary counterexamples
to the claim. With the aim of unifying the treatment of as many special
cases of physical interest as possible, it is then necessary to consider $%
\mathcal{M}$ as an abstract manifold prescribing only the minimal
geometrical properties necessary to build up the essential objects which are
useful for constructing the mechanics of complex bodies. In this case the
existence of the microstress $\mathcal{S}\left( x\right) \in Hom(\mathbb{%
\hat{R}}^{d},T_{\nu \left( x\right) }^{\ast }\mathcal{M})$ can be assumed a
priori (an intrinsic representation of $\mathcal{S}$ in terms of measures is
presented in \cite{S} where the primary object considered is the inner
power, instead of the external one).

Without investigating further on the question, I assume here that $%
\boldsymbol{\tau }$ depends linearly on the normal at $x$ to $\partial 
\mathfrak{b}$ so that the natural expression of the external power of all
actions over the generic part $\mathfrak{b}$ along $y$ and $\nu $ is then%
\begin{equation*}
\mathcal{P}_{\mathfrak{b}}^{ext}\left( \dot{y},\dot{\nu}\right) :=\int_{%
\mathfrak{b}}\left( b\cdot \dot{y}+\beta \cdot \dot{\nu}\right) \text{ }%
dx+\int_{\partial \mathfrak{b}}\left( Pn\cdot \dot{y}+\mathcal{S}n\cdot \dot{%
\nu}\right) \text{ }d\mathcal{H}^{d-1}.
\end{equation*}

A crucial axiom is the requirement that the power $\mathcal{P}_{\mathfrak{b}%
}^{ext}$\ is invariant under isometric changes in observers, that is under
the action of the semi-direct product $\mathbb{\hat{R}}^{d}\ltimes SO\left(
d\right) $ over the ambient space and the action of elements of $G$ over $%
\mathcal{M}$ induced by the homomorphism $h$ introduced above (see also
additional remarks in \cite{M02}, \cite{M08}, \cite{MM}). Here, a weaker
axiom is used, namely invariance of the power under the sole action of $%
SO\left( d\right) $ in space and the corresponding action of $G$ over $%
\mathcal{M}$ through $h$. It is thus required that observers differing only
by a proper rotation evaluate the same power.

\ \ \ \ \ \ \ \ \ \ \ \ \ \ \ \ \ \ 

\begin{axiom}
($SO\left( d\right) $ invariance) At mechanical equilibrium the external
power of all actions on any part of $\mathcal{B}$ is invariant under
rotational semi-classical changes in observers, namely%
\begin{equation}
\mathcal{P}_{\mathfrak{b}}^{ext}\left( \dot{y}^{\ast },\dot{\nu}^{\ast
}\right) =\mathcal{P}_{\mathfrak{b}}^{ext}\left( \dot{y},\dot{\nu}\right)
\label{A}
\end{equation}%
for any choice of the rotational velocity $q\left( t\right) \in \mathbb{\hat{%
R}}^{d}$ and any $\mathfrak{b}\in \mathfrak{P}\left( \mathcal{B}\right) $.
\end{axiom}

\ \ \ \ \ \ \ \ \ \ \ 

\begin{theorem}
(i) If for any $\mathfrak{b}$\ the vector fields $x\mapsto Pn$ and $x\mapsto 
\mathcal{A}^{\ast }\mathcal{S}n$ are defined over $\partial \mathfrak{b}$
and are integrable there, the integral balances of actions on $\mathfrak{b}$
\ hold: 
\begin{equation}
\int_{\mathfrak{b}}b\text{ }dx+\int_{\partial \mathfrak{b}}Pn\text{ }d%
\mathcal{H}^{d-1}=0,  \label{IntFor}
\end{equation}%
\begin{equation}
\int_{\mathfrak{b}}\left( \left( y-y_{0}\right) \wedge b+\mathcal{A}^{\ast
}\beta \right) \text{ }dx+\int_{\partial \mathfrak{b}}\left( \left(
y-y_{0}\right) \wedge Pn+\mathcal{A}^{\ast }\mathcal{S}n\right) \text{ }d%
\mathcal{H}^{d-1}=0.  \label{IntMomen}
\end{equation}%
(ii) Moreover, if the tensor fields $x\mapsto P$ and $x\mapsto \mathcal{S}$
are of class $C^{1}\left( \mathcal{B}_{0}\right) \cap C^{0}\mathcal{(\bar{B}}%
_{0})$ then 
\begin{equation}
DivP+b=0  \label{Cau}
\end{equation}%
and there exist a covector field $x\mapsto z\in T_{\nu \left( x\right) }%
\mathcal{M}$ such that 
\begin{equation}
skw\left( PF^{\ast }\right) =\mathsf{e}\left( \mathcal{A}^{\ast }z+\left( D%
\mathcal{A}^{\ast }\right) \mathcal{S}\right)  \label{SKW}
\end{equation}%
and 
\begin{equation}
Div\mathcal{S}-z+\beta =0,  \label{Cap}
\end{equation}%
with $z=z_{1}+z_{2}$, $z_{2}\in Ker\mathcal{A}^{\ast }$. (iii) If the rate
fields $\left( x,t\right) \longmapsto \dot{y}\left( x,t\right) \in \mathbb{%
\hat{R}}^{d}$ and $\left( x,t\right) \longmapsto \dot{\nu}\left( x,t\right)
\in T_{\nu \left( x\right) }\mathcal{M}$ are differentiable in space, the
local balances imply%
\begin{equation}
\mathcal{P}_{\mathfrak{b}}^{ext}\left( \dot{y},\dot{\nu}\right) =\mathcal{P}%
_{\mathfrak{b}}^{int}\left( \dot{y},\dot{\nu}\right)  \label{PLV}
\end{equation}%
where%
\begin{equation*}
\mathcal{P}_{\mathfrak{b}}^{int}\left( \dot{y},\dot{\nu}\right) :=\int_{%
\mathfrak{b}}(P\cdot \dot{F}+z\cdot \dot{\nu}+\mathcal{S}\cdot \dot{N})\text{
}dx.
\end{equation*}
\end{theorem}

Above \textsf{e} is Ricci's alternating index. $\mathcal{P}_{\mathfrak{b}%
}^{int}\left( \dot{y},\dot{\nu}\right) $\ is called an inner (or internal)
power.

\begin{proof}
The immediate consequence of the axiom of $SO\left( d\right) $ invariance is
the integral balance of moments (\ref{IntMomen}) obtained by using (\ref{CO1}%
) and (\ref{CO2}) in (\ref{A}). In (\ref{CO1}) the point $y_{0}$ is
arbitrary. As a consequence, by taking an arbitrary vector $w\in \mathbb{%
\hat{R}}^{d}$, one can substitute $y_{0}$ with $y_{0}+w$ in (\ref{IntMomen}%
). Such a substitution corresponds to a simple shift of the centre of the
rotation of one observer with respect to the other. By subtracting (\ref%
{IntMomen}) from its counterpart calculated at $y_{0}+w$, one then gets%
\begin{equation*}
w\cdot (\int_{\mathfrak{b}}b\text{ }dx+\int_{\partial \mathfrak{b}}Pn\text{ }%
d\mathcal{H}^{d-1})=0,
\end{equation*}%
which corresponds to (\ref{IntFor}) as a consequence of the arbitrariness of 
$w$. Note that the substitution $y_{0}\longrightarrow y_{0}+w$ is possible
due to the linear structure of $\mathbb{\hat{R}}^{d}$, a structure that is
in general not available over the manifold of substructural shapes $\mathcal{%
M}$. Under the regularity hypotheses above, the local balance (\ref{Cau})
follows as usual by exploiting Gauss theorem and the arbitrariness of $%
\mathfrak{b}$. The same localization procedure applied to the integral
balance (\ref{IntMomen}) and the validity of (\ref{Cau}) imply the local
balance%
\begin{equation*}
\mathsf{e}PF^{\ast }-\left( D\mathcal{A}^{\ast }\right) \mathcal{S}=\mathcal{%
A}^{\ast }\left( Div\mathcal{S}+\beta \right) .
\end{equation*}%
Since $\mathcal{A}^{\ast }\left( \nu \right) \in Hom(T_{v}^{\ast }\mathcal{M}%
,\mathbb{\hat{R}}^{d})$, two information are available from this equation:
(1) At each $\nu \in \mathcal{M}$ the difference $\mathsf{e}PF^{\ast
}-\left( D\mathcal{A}^{\ast }\right) \mathcal{S}$ is the image in $\mathbb{%
\hat{R}}^{d}$ of a covector in $T_{v}^{\ast }\mathcal{M}$, let say $z$. (2)
Such a covector is just equal to $Div\mathcal{S}+\beta $. The equation (\ref%
{PLV}) follows by direct calculation under the validity of the pointwise
balances (\ref{Cau}) and (\ref{Cap}).
\end{proof}

Of course, the balance equations above include the dynamic case because the
bulk actions can be decomposed additively in their inertial and non-inertial
parts, the latter being identified by requiring that their power balances
the rate of change of the kinetic energy. In this procedure it is assumed
that the energy is the sum of macroscopic and substructural contributions.
For the sake of conciseness the topic is not developed here (see \cite{M08}
for the details).

Theorem 1 is the same that can be obtained by imposing as an axiom the
invariance of the external power under isometric semi-classical changes in
observers mentioned earlier (see \cite{M02}). Such an equivalence underlines
that the integral balance of standard forces is a peculiar consequence of
the `rigid' structure of $\mathbb{\hat{R}}^{d}$ (its linear structure) and
is associated with one of the Killing fields of the metric in the ambient
space (see \cite{S-book} for the analogous observation in the case of simple
bodies). The same property is not available over $\mathcal{M}$ straight
away. In fact, it has been assumed here that the manifold of substructural
shapes is abstract so that it does not coincide with a linear space in
general. For this reason the totals of the substructural actions are not
defined a priori, as mentioned above in discussing the possible path toward
the proof of the existence of the microstress $\mathcal{S}$. Consider, for
example, the field $x\longmapsto \beta \left( x\right) \in T_{\nu \left(
x\right) }^{\ast }\mathcal{M}$ that is $\beta :\mathcal{B}\longrightarrow
T^{\ast }\mathcal{M}$. The target space $T^{\ast }\mathcal{M}$\ is not a
linear space so that the integral of $\beta $ on any part $\mathfrak{b}$\ of 
$\mathcal{B}$\ is not defined unless $\mathcal{M}$ itself is a linear space.
Analogous remarks hold for the fields $x\longmapsto z\left( x\right) \in
T_{\nu \left( x\right) }^{\ast }\mathcal{M}$ and $x\longmapsto \left( 
\mathcal{S}n\right) \left( x\right) \in T_{\nu \left( x\right) }^{\ast }%
\mathcal{M}$. As a consequence, not only an integral balance of
substructural actions does not follow from the requirement of invariance of
the power under changes in observers but it is even not defined. Moreover,
the fact that $\beta $\ and $\mathcal{S}$\ appear only in the balance of
moments does not means that they represent (micro) couples because in the
integral balance of moments they are multiplied by the the formal adjoint of 
$\mathcal{A}$ which maps at each $\nu $ elements of $T_{\nu }\mathcal{M}$
onto elements of $\mathbb{\hat{R}}^{d}$.

Various alternative paths can be followed to get pointwise balances of
actions. The comments below apply to them.

\begin{enumerate}
\item One could postulate the integral balances of standard and
substructural actions as first principles. However, as pointed out above,
such a point of view can be adopted only in the (very) special case in which 
$\mathcal{M}$ is a linear space. When it is not the case, the balance of
substructural actions cannot be postulated, because the integrals appearing
are not defined.

\item One could adopt the virtual power procedure proposed in \cite{Ger} for
affine bodies, by postulating in fact the weak form of the balance
equations. In this case one must postulate not only the expression of the
external power but also the internal power, the power of the inner actions.
In this way one should postulate the existence of the inner self-action $z$.
In contrast, in Theorem 1 the existence of $z$ is proven.

\item One could adopt the point of view by Green, Rivlin and Naghdi along
the path indicated by Marsden's and Hughes's theorem in \cite{MH} by
postulating the expression of the first principle of thermodynamics (a point
of view exploited in \cite{Quart} with reference to isometric changes in
observers). In this case, however, one is forced not only to postulate the
existence of the energy but also to prescribe its functional dependence on
the state variables, in contrast with the minimalist approach followed in
Theorem 1. Such a point of view is however one of the manners useful to
prove the covariance of the pointwise balance of actions (that is the
invariance under the action of the entire group of diffeomorphisms on the
ambient space and the action of $G$ on $\mathcal{M}$). The other ways are
given by the exploitation of Noether theorem and/or d'Alembert-Lagrange type
principles in presence of viscous-type dissipation at macroscopic and/or
substructural level (see \cite{dFM} and \cite{M08} for the relevant
results). A requirement of covariance allows one to eliminate the
indetermination given by $z_{2}\in Ker\mathcal{A}^{\ast }$: in this case $%
z_{2}$ vanishes identically.
\end{enumerate}

When $\nu $ represents only a generic property of the substructures not
associated with their geometry in space, changes of frame in the ambient
space and over $\mathcal{M}$ can be considered disconnected. By imposing
invariance of the external power with respect to isometric semi-classical
changes in observers, one gets two distinct integral balances of moments:%
\begin{equation*}
\int_{\mathfrak{b}}\left( y-y_{0}\right) \wedge b\text{ }dx+\int_{\partial 
\mathfrak{b}}\left( y-y_{0}\right) \wedge Pn\text{ }d\mathcal{H}^{d-1}=0
\end{equation*}%
and%
\begin{equation*}
\int_{\mathfrak{b}}\mathcal{A}^{\ast }\beta \text{ }dx+\int_{\partial 
\mathfrak{b}}\mathcal{A}^{\ast }\mathcal{S}n\text{ }d\mathcal{H}^{d-1}=0.
\end{equation*}%
Theorem 1 can be rewritten. The sole difference in this case is that \ref%
{SKW} splits in the two equations%
\begin{equation}
skw\left( PF^{\ast }\right) =0,\text{ \ \ }skw\left( \mathcal{A}^{\ast
}z+\left( D\mathcal{A}^{\ast }\right) \mathcal{S}\right) =0.  \label{Disc}
\end{equation}

\section{The energetic scenario and the existence of ground states}

\subsection{A priori restrictions on constitutive structures.}

After describing the morphology of the generic material element and
representing the actions along a motion, the local energetic scenario must
be specified.

At the macroscopic scale, since deformation is accounted for, each material
element (considered as a whole) is assumed in energetic contact with the
neighboring fellows. The consequent interactions are standard tensions.

At the scale of the substructure, i.e. within each material element, some
alternatives are possible (see additional remarks in \cite{QuadMat}). They
are classified under suggestion of the common path followed in statistical
physics.

\begin{enumerate}
\item The generic material element is a closed system with respect to its
substructure: there is no migration of substructures out of the material
element, and the substructure itself does not interact energetically with
the neighboring fellows.

\item The substructure of the generic material element is in energetic
contact with the substructures of the neighboring elements. No migration
occur.

\item The material element is an open system: both energetic contact and
migration of substructures are possible.
\end{enumerate}

Here the attention is focused on case 2 mainly. Remarks are added on case 1.
Case 3 is not touched here for the sake of brevity (see \cite{M05} for
relevant developments).

The procedure to establish a priori constitutive restrictions is the
standard one, based on the Clausius-Duhem inequality which is written here
in isothermal form as a mechanical dissipation inequality. It prescribes that%
\begin{equation}
\frac{d}{dt}\Psi \left( \mathfrak{b}\right) -\mathcal{P}_{\mathfrak{b}%
}^{ext}\left( \dot{y},\dot{\nu}\right) \leq 0,  \label{MDI}
\end{equation}%
for any choice of the rate fields. $\Psi \left( \mathfrak{b}\right) $ is the
total \emph{free energy} of $\mathfrak{b}$ along $\left( x,t\right)
\longmapsto \left( y\left( x,t\right) ,\nu \left( x,t\right) \right) $. The
standard assumption is that $\Psi \left( \mathfrak{b}\right) $ is absolutely
continuous with respect to the volume measure so that there is a density $%
\psi $ such that%
\begin{equation*}
\Psi \left( \mathfrak{b}\right) =\int_{\mathfrak{b}}\psi \text{ }dx.
\end{equation*}%
The constitutive dependence on the state variables must then be assigned not
only for $\psi $ but also for the stress measures (namely $P$, $z$ and $%
\mathcal{S}$). Simple assumptions are as follows:%
\begin{equation*}
\psi =\psi \left( F,\nu ,N\right) ,\text{ \ \ }P=P\left( F,\nu ,N\right) ,%
\text{ \ \ }z=z\left( F,\nu ,N\right) ,\text{ \ \ }\mathcal{S}=\mathcal{S}%
\left( F,\nu ,N\right) .
\end{equation*}%
If $\psi $ admits partial derivatives with respect to its entries, the
arbitrariness of $\mathfrak{b}$\ and equation (\ref{PLV}) imply the local
dissipation inequality%
\begin{equation}
\left( \partial _{F}\psi -P\right) \cdot \dot{F}+\left( \partial _{\nu }\psi
-z\right) \cdot \dot{\nu}+\left( \partial _{N}\psi -\mathcal{S}\right) \cdot 
\dot{N}\leq 0.  \label{LDI}
\end{equation}%
The possibility to choose arbitrarily the rate in (\ref{LDI}) from any given
state $\left( F,\nu ,N\right) $ implies the classical relations (see also 
\cite{C89}) 
\begin{equation}
P=\partial _{F}\psi \left( F,\nu ,N\right) ,\text{ \ \ \ }z=\partial _{\nu
}\psi \left( F,\nu ,N\right) ,\text{ \ \ \ }\mathcal{S}=\partial _{N}\psi
\left( F,\nu ,N\right) .  \label{CR}
\end{equation}

The mechanical dissipation inequality (\ref{MDI}) forbids the dependence of $%
\psi $ on the rate of the fields involved. In fact, if $\psi $ would depend
on (let say) $\dot{\nu}$, in the reduced version of the mechanical
dissipation inequality a term of the type $\partial _{\dot{\nu}}\psi \left(
F,\nu ,N,\dot{\nu}\right) \cdot \ddot{\nu}$ would appear with no
correspondence in the structure of the internal power $\mathcal{P}_{%
\mathfrak{b}}^{int}\left( \dot{y},\dot{\nu}\right) $ where no action
developing power in $\ddot{\nu}$ is presented. The arbitrariness of $\ddot{%
\nu}$\ would imply then $\partial _{\dot{\nu}}\psi =0$. In contrast, $P$, $z$
and $\mathcal{S}$ may depend on the rates of the state variables when \emph{%
viscous-like effects occur }at various scales. The dependence on the the
rate of the state variables is compatible with the mechanical dissipation
inequality \ref{MDI}, provided that one assumes the validity of an additive
decomposition of $P$, $z$ and $\mathcal{S}$ into conservative and
dissipative parts. By indicating the triple $\left( F,\nu ,N\right) $ by $%
\varsigma $, one then presumes that%
\begin{equation*}
P=P^{c}\left( \varsigma \right) +P^{d}\left( \varsigma ,\dot{\varsigma}%
\right) ,\text{ \ \ }z=z^{c}\left( \varsigma \right) +z^{d}\left( \varsigma ,%
\dot{\varsigma}\right) ,\text{ \ \ }\mathcal{S}=\mathcal{S}^{c}\left(
\varsigma \right) +\mathcal{S}^{d}\left( \varsigma ,\dot{\varsigma}\right) .
\end{equation*}%
Such decompositions must be supplemented by the assumption that the
conservative components are determined by the free energy. The use of (\ref%
{MDI}) implies once more the relations (\ref{CR}) for the conservative
addenda and the reduced dissipation inequality%
\begin{equation}
P^{d}\cdot \dot{F}+z^{d}\cdot \dot{\nu}+\mathcal{S}^{d}\cdot \dot{N}\geq 0.
\label{RD1}
\end{equation}%
Consequently, $P^{d}$, $z^{d}$ and $\mathcal{S}^{d}$\ are linear in $\dot{F}$%
, $\dot{\nu}$ and $\dot{N}$. Additional assumptions on the structure of the
dissipation can be made.

\begin{enumerate}
\item One may presume that strong dissipation conditions are satisfied a
priori (that is independently of \ref{MDI}), namely%
\begin{equation*}
P^{d}\cdot \dot{F}\geq 0,\text{ \ \ }z^{d}\cdot \dot{\nu}\geq 0,\text{ \ }%
\mathcal{S}^{d}\cdot \dot{N}\geq 0.
\end{equation*}%
Then one may write%
\begin{equation*}
P^{d}=a_{P}\dot{F},\text{ \ \ }z^{d}=a_{z}\dot{\nu},\text{ \ \ }\mathcal{S}%
^{d}=a_{S}\dot{N},
\end{equation*}%
with $a_{P}$, $a_{z}$ and $a_{S}$ positive definite (scalar valued) state
functions.

\item One could consider a strong condition for the macroscopic dissipation,
namely 
\begin{equation*}
P^{d}\cdot \dot{F}\geq 0
\end{equation*}%
and a weaker dissipation condition for the substructure:%
\begin{equation*}
z^{d}\cdot \dot{\nu}+\mathcal{S}^{d}\cdot \dot{N}\geq 0.
\end{equation*}%
In this case, $P^{d}$ is equal to $a_{P}\dot{F}$ while $z^{d}$ and $\mathcal{%
S}^{d}$ are linear functions of $\dot{\nu}$ and $\dot{N}$.

\item Other conditions can be presumed to hold. They may describe different
viscous-like effects. Dissipative effects of plastic-like type can be
accounted for. The standard plasticity theory and its strain gradient
version fall within this scheme, when one identifies $\nu $ with the plastic
strain. In all cases of plastic-like behavior a flow condition in terms of
the subdifferential of some admissible region in the state space must be
involved (here the existence of a dissipation pseudo-potential is assumed).
Energetic solutions to the resulting \ evolutionary problem can be obtained
under appropriate hypotheses (relevant analytical tools can be found in \cite%
{Mi}, \cite{MiF}).
\end{enumerate}

One may ask what is the relation with standard internal variable schemes. In
their initial formulation, such schemes have been proposed with the aim of
describing the removal from thermodynamical equilibrium (see \cite{DM}, \cite%
{Gy}). In this (historical) sense internal variables are by definition not
observable and play a parametric role at equilibrium, in contrast with the
approach proposed here. The derivatives of the energy with respect to the
internal variables and their derivatives are not considered true
interactions, rather they are thermodynamic affinities (see once more \cite%
{Mi}, \cite{MiF}). They do not appear in the expression of the external
power. In contrast, I consider $\nu $ as an observable quantity the
variations of which contribute to the equilibrium by means of true
interactions. This is the reason for which I call $\nu $\ morphological
descriptor rather than internal variable. Connections are possible between
the internal variable scheme and the multifield scheme that I discuss here.
Assumptions should be necessary in order to avoid to render the comparison
only formal.

\subsection{Ground states}

Consider a complex body displaying a pure conservative behavior. In this
case one may identify the free energy with the elastic one. In absence of
inertia, the energy of the whole body is then%
\begin{equation*}
\mathcal{E}\left( y,\nu \right) :=\int_{\mathcal{B}}e\left( x,y\left(
x\right) ,F\left( x\right) ,\nu \left( x\right) ,N\left( x\right) \right) 
\text{ }dx,
\end{equation*}%
where the density $e$\ is the difference between the elastic energy $e^{i}$
and the potential of body forces $e_{1}^{e}+e_{2}^{e}$, the latter being
decomposed in the part associated with standard gravitational forces $%
(e_{1}^{e})$ and the potential of possible external fields acting directly
over the substructure $\left( e_{2}^{e}\right) $, namely $e=e^{i}\left( x,F%
\mathbf{,}\nu \mathbf{,}N\right) -\left( e_{1}^{e}\left( u\right)
+e_{2}^{e}\left( \nu \right) \right) $, with $e^{i}$ the elastic energy.

A pair of fields $\left( u,\nu \right) $ satisfying the variational
principle 
\begin{equation*}
\min_{y,\nu }\mathcal{E}\left( y,\nu \right)
\end{equation*}%
is called \emph{ground state}. Conditions for the existence of ground states
follow constitutive assumptions on (\emph{i}) the nature of the functional
classes in which one places $y$ and $\nu $, (\emph{ii}) the `structural'
properties of $e$.

Let $y:\mathcal{B}\rightarrow \mathbb{\hat{R}}^{d}$ be a Sobolev map, namely
an element of $W^{1,1}(\mathcal{B},\mathbb{\hat{R}}^{d})$. Denote first by $%
M\left( F\right) $ the $d-$vector in $\mathbb{\hat{R}}^{2d}$\ collecting all
the minors of $F$ (i.e. of $Dy$). $M\left( F\right) $ is then an element of $%
\Lambda _{d}(\mathbb{R}^{d}\times \mathbb{\hat{R}}^{d})$.

It is possible to construct the $d-$\emph{current integration} $G_{y}$ over
the graph of $y$. Precisely, $G_{y}$ is the linear functional on smooth $d-$%
forms $\omega $ with compact support in $\mathcal{B}\times \mathbb{\hat{R}}%
^{d}$ defined by%
\begin{equation*}
G_{y}:=\int_{\mathcal{B}}\left\langle \omega \left( x,y\left( x\right)
\right) ,M\left( Dy\left( x\right) \right) \right\rangle \text{ }dx\mathbf{.}
\end{equation*}%
The \emph{boundary current} associated with $G_{y}$\ is indicated by $%
\partial G_{y}$\ and defined by $\partial G_{y}\left( \omega \right)
:=G_{u}\left( d\omega \right) $,\ $\omega \in \mathcal{D}^{d-1}(\mathcal{B}%
\times \mathbb{\hat{R}}^{d})$ with $\mathcal{D}^{d-1}(\mathcal{B}\times 
\mathbb{\hat{R}}^{d})$ the space of $(d-1)-$forms with compact support in $%
\mathcal{B}\times \mathbb{\hat{R}}^{d}$ (details on the nature and the
properties of Cartesian currents can be found in \cite{GMS}).

The functional spaces in which the existence of minima is investigated must
be specified. Their choice has constitutive nature.

\begin{enumerate}
\item The macroscopic deformation $y$ is assumed to be a \emph{weak
diffeomorphism} (in symbols $y\in dif^{1,1}(\mathcal{B},\mathbb{\hat{R}}%
^{d}) $). In fact, $y$ is considered a $W^{1,1}(\mathcal{B},\mathbb{\hat{R}}%
^{d})$ map such that (\emph{i}) $\left\vert M\left( Dy\right) \right\vert
\in L^{1}\left( \mathcal{B}\right) $, (\emph{ii}) $\partial G_{y}=0$ on $%
\mathcal{D}^{d-1}(\mathcal{B}\times \mathbb{\hat{R}}^{d})$, (\emph{iii}) $%
\det Dy\left( x\right) >0$ for almost every $x\in \mathcal{B}$, (\emph{iv})
for any $f\in C_{c}^{\infty }(\mathcal{B}\times \mathbb{\hat{R}}^{d})$%
\begin{equation*}
\int_{\mathcal{B}}f\left( x\mathbf{,}y\left( x\right) \right) \det Dy\left(
x\right) \text{ }dx\leq \int_{\mathbb{\hat{R}}^{d}}\sup_{x\mathbf{\in }%
\mathcal{B}}f\left( x,w\right) \text{ }dw\mathbf{.}
\end{equation*}%
In particular, the subspace%
\begin{equation*}
dif^{r,1}(\mathcal{B},\mathbb{\hat{R}}^{d}):=\left\{ y\in dif^{1,1}(\mathcal{%
B},\mathbb{\hat{R}}^{d})|\text{ }\left\vert M\left( Dy\right) \right\vert
\in L^{r}\left( \mathcal{B}\right) \right\} ,
\end{equation*}%
for some $r>1$, is of special interest below.

\item It is assumed that (\emph{i}) $\mathcal{M}$ has Riemannian structure
with (at least) $C^{1}-$metric $g_{\mathcal{M}}$, and (\emph{ii}) covariant
derivatives are explicitly calculated by making use of the natural
Levi-Civita connection. The $C^{1}-$Riemannian structure implies that $%
\mathcal{M}$ can be isometrically embedded in a linear space isomorphic to $%
\mathbb{R}^{M}$ (for some $M$) by Nash theorem: it is then considered as a
closed submanifold of $\mathbb{R}^{M}$. It is then assumed that $\nu \in
W^{1,s}\left( \mathcal{B}_{0},\mathcal{M}\right) $, $s>1$, with 
\begin{equation*}
W^{1,s}\left( \mathcal{B},\mathcal{M}\right) :=\left\{ \nu \in W^{1,s}\left( 
\mathcal{B},\mathbb{R}^{M}\right) \text{ }|\text{ }\nu \left( x\right) \in 
\mathcal{M}\text{ for a.e. }x\right\} .
\end{equation*}%
The energy functional $\mathcal{E}$ is then extended to 
\begin{equation*}
\mathcal{W}_{r,s}:=dif^{r,1}(\mathcal{B},\mathbb{\hat{R}}^{d})\times
W^{1,s}\left( \mathcal{B},\mathcal{M}\right) .
\end{equation*}
\end{enumerate}

Assumptions on the structural properties of the energy density $e$ must also
be specified. $e$ cannot be convex in $F$ for the standard objectivity
argument but it can be convex in $N$.

\begin{itemize}
\item $e$ is assumed to be polyconvex in $F$ and convex in $N$. There exists
a Borel function $Pe:\mathcal{B}\times \mathbb{\hat{R}}^{d}\times \mathcal{M}%
\times \Lambda _{d}(\mathbb{R}^{d}\times \mathbb{\hat{R}}^{d})\times
M_{M\times d}\rightarrow {\bar{\mathbb{R}}}^{+}$, with values $Pe\left(
x,y,\nu ,\xi ,N\right) $, which is (\emph{i}) l. s. c. in $\left( y,\nu ,\xi
,N\right) $ for a.e. $x\in \mathcal{B}$, (\emph{ii}) convex in $\left( \xi
,N\right) $\ for any $\left( x,y,\nu \right) $, (\emph{iii}) and also such
that $Pe\left( x,y,\nu ,M\left( F\right) \mathbf{,}N\right) =e\left( x,y,\nu
,F\mathbf{,}N\right) $ for any $\left( x,y,\nu ,F\mathbf{,}N\right) $ with $%
\det F>0$.

\item By assumption $e$ satisfies the growth condition 
\begin{equation*}
e\left( x,y,\nu ,F\mathbf{,}N\right) \geq C_{1}\left( \left\vert M\left(
F\right) \right\vert ^{r}+\left\vert N\right\vert ^{s}\right) +\vartheta
\left( \det F\right)
\end{equation*}%
for any $\left( x,y,\nu ,F\mathbf{,}N\right) $ with $\det F>0$, $r,s>1$ and $%
C_{1}>0$ constants, and $\vartheta :\left( 0,+\infty \right) \rightarrow 
\mathbb{R}^{+}$ a convex function such that $\vartheta \left( t\right)
\rightarrow +\infty $ as $t\rightarrow 0^{+}$.
\end{itemize}

\begin{theorem}
(\cite{MM}) The functional $\mathcal{E}$ achieves the minimum value in the
classes 
\begin{equation*}
\mathcal{W}_{r,s}^{d}:=\left\{ \left( y,\nu \right) \in \mathcal{W}%
_{r,s}|y=y_{0}\text{ on }\partial \mathcal{B}_{y},\nu =\nu _{0}\text{ on }%
\partial \mathcal{B}_{\nu }\right\}
\end{equation*}%
and%
\begin{equation*}
\mathcal{W}_{r,s}^{c}:=\left\{ \left( y,\nu \right) \in \mathcal{W}_{r,s}%
\text{ }|\text{ }\partial G_{y}=\partial G_{y_{0}}\text{ on }\mathcal{D}^{2}(%
\mathbb{R}^{d}\times \mathbb{\hat{R}}^{d}),\nu =\nu _{0}\text{ on }\partial 
\mathcal{B}_{\nu }\right\} .
\end{equation*}
\end{theorem}

In the theorem above $\partial \mathcal{B}_{y}$ and $\partial \mathcal{B}%
_{\nu }$ are the portions of the boundary $\partial \mathcal{B}$\ where
boundary data are assigned in terms of $y$ and $\nu $ respectively. Details
and comments on the physical consequences of the assumptions above can be
found in \cite{MM}.

\section{Notes and complements}

The first variation of the energy functional $\mathcal{E}\left( y,\nu
\right) $\ along $C^{1}$ minimizers allows one to obtain the balances of
standard and substructural actions (\ref{Cau}) and (\ref{Cap}). The
condition (\ref{SKW}) is a consequence of a requirement of objectivity for
the elastic energy $e^{i}$. Remarks leading to (\ref{Disc}) also apply. In
addition, horizontal variations can be made by altering the reference place
by means of the diffeomorphism $\Phi _{\varepsilon }\left( x\right)
:=x+\varepsilon \phi \left( x\right) $, $\phi \in C_{0}^{1}\left( \mathcal{B}%
,\mathbb{R}^{3}\right) $, $\varepsilon $ a real parameter. $\Phi
_{\varepsilon }\left( x\right) $ leaves unchanged the boundary $\partial 
\mathcal{B}$ for $\varepsilon $ sufficiently small. In fact, horizontal
variations can be considered as a sort of relabeling of the reference place.
One then defines $y_{\varepsilon }\left( x\right) :=y\left( \Phi
_{\varepsilon }^{-1}\left( x\right) \right) $ and $\nu _{\varepsilon }\left(
x\right) :=\nu \left( \Phi _{\varepsilon }^{-1}\left( x\right) \right) $,
and obtains a mapping $\varepsilon \longmapsto \mathcal{E}\left(
y_{\varepsilon },\nu _{\varepsilon }\right) $. In case of appropriate
smoothness, differentiation with respect to $\varepsilon $ gives rise to the
configurational balance%
\begin{equation}
Div\mathbb{P}+\partial _{x}e=0.  \label{Conf}
\end{equation}%
where $\mathbb{P}=e^{i}I-F^{\ast }P-N^{\ast }\mathcal{S}\in Aut\left( 
\mathbb{R}^{d}\right) $ is the extended Hamilton-Eshelby tensor in the
mechanics of complex bodies (see \cite{M02}). In the smooth case (\ref{Conf}%
) is essentially the pull back in $\mathcal{B}$ of the balance of standard
forces (\ref{Cau}).

Different is the stage when the variation is calculated on non-smooth
minimizers, namely on $\mathcal{W}_{r,s}^{d}$ where it is sure by Theorem 2
that ground states exist. The main difficulty is that Sobolev maps may not
admit tangential derivatives so that the balance of standard forces in terms
of first Piola-Kirchhoff stress (\ref{Cau}) cannot be derived. One may
compute horizontal variations, variations over $\mathcal{M}$ and the
variation of the actual shape $\mathcal{B}_{a}$. I summarize here the
relevant results (see for details \cite{MM}).

\begin{enumerate}
\item Under appropriate growth conditions for the polyconvex extension of
the energy and its derivatives with respect to $x$, $M\left( F\right) $ and $%
N$, one proves that (\emph{i}) $F^{\ast }P$ and $N^{\ast }\mathcal{S}$
belong to $L^{1}\left( \mathcal{B}\right) $ and (\emph{ii}) the balance (\ref%
{Conf}) holds in terms of distributions.

\item An additional assumption on the growth of $\left\vert \partial
_{y}Pe\right\vert $ implies that (\emph{i}) $\sigma :=\left( \det F\right)
^{-1}PF^{\ast }\in L_{loc}^{1}(\tilde{y}\left( \mathcal{B}_{0}\right) ,%
\mathbb{\hat{R}}^{3}\otimes \mathbb{\hat{R}}^{3})$, with $x\longmapsto 
\tilde{y}\left( x\right) $ the Lusin representative of $y$ and (\emph{ii})
the balance of standard forces hold in distributional sense in terms of the
Cauchy stress $\sigma $.

\item Variations over $\mathcal{M}$, obtained by means of smooth curves $%
\varepsilon \rightarrow \bar{\varphi}_{\varepsilon }\in Aut\left( \mathcal{M}%
\right) $, $\bar{\varphi}\in C^{1}\left( \mathcal{M}\right) $, and an
additional assumption on the growth of $\left\vert \partial _{\nu
}Pe\right\vert $, allow one to show that $\mathcal{S}\in L^{1}\left( 
\mathcal{B}_{0},\mathbb{R}^{3\ast }\otimes T^{\ast }\mathcal{M}\right) $ and
(\emph{ii}) the balance of substructural actions (\ref{Cap}) holds in the
sense of distributions. Such a result excludes the interpretation\
(suggested by some author) of the balance of substructural actions as a sort
of balance of configurational actions, unless the word configurational is
used in a sense a little bit far from the current one.
\end{enumerate}

The remarks in this section contribute to the current debate on the nature
of configurational actions.

\ \ \ \ \ \ \ \ \ \ \ \ \ \ \ \ \ \ \ \ \ \ \ 

\textbf{Acknowledgements}. This paper is the extended version of a talk that
I delivered at Zurich in July 2007, during the ICIAM. The support of the
Italian National Group for Mathematical Physics, GNFM-INDAM, is
acknowledged. I whish to thank also the "Centro di Ricerca Matematica Ennio
De Giorgi" of the "Scuola Normale Superiore" at Pisa for providing an
appropriate environment for scientific interactions.

\end{document}